# A definition of the coupled-product for multivariate coupled-exponentials

Kenric P. Nelson*

Raytheon Company, 50 Apple Hill Dr., Tewksbury, MA 01876

## Abstract

The coupled-product and coupled-exponential of the generalized calculus of nonextensive statistical mechanics are defined for multivariate functions. The nonlinear statistical coupling is indexed such that $\kappa_d = \kappa / 1 + d\kappa$, where $d$ is the dimension of the argument of the multivariate coupled-exponential. The coupled-Gaussian distribution is defined such that the argument of the coupled-exponential depends on the coupled-moments but not the coupling parameter. The multivariate version of the coupled-product is defined such that the output dimensions are the sum of the input dimensions. This enables construction of the multivariate coupled-Gaussian from univariate coupled-Gaussians. The resulting construction forms a model of coupling between distributions, generalizing the product of independent Gaussians.

## 1. Introduction

A definition for a generalization of the exponential family based on the degree of nonlinear statistical coupling is provided. The purpose of the definition is to facilitate development of a generalized calculus which will enable analytic reasoning for nonlinear systems, while retaining as much of the structure of the exponential family and its connection to linear systems. The definition builds upon the generalized calculus of nonextensive entropy [1, 2], but incorporates several new features. First, the proposed generalized calculus is defined using the nonlinear statistical coupling $\kappa = 1 - q$ [3], a translation of the Tsallis entropy index [4]. This translation has also been utilized by for example [1, 5-7]. Second, the coupling terms for the arguments and the exponent have a non-equal functional relationship [8]. Third, the two coupling terms are defined such that the argument of the coupled-exponential and coupled-Gaussian distribution only depend on the generalized parameters of the distribution. Fourth, the dependence on the dimensions of the argument is separated from the coupling. The purpose of these modifications is to isolate the physical effects of the nonlinear coupling and to define the coupled-product consistent with the change in dimensions of a multivariate coupled-exponential [8] from the operands to the resultant.

---

* Corresponding Author: Phone: 603-508-9827 Email: kenric_p_nelson@raytheon.com

## 2. Definitions of the coupled-exponential and related algebra

The concept of nonlinear statistical coupling is based on two mathematical features of the generalized calculus for nonextensive statistical mechanics. Firstly, the coupled-sum $x \oplus_\kappa y = x + y + \kappa xy$ modifies addition to include a nonlinear term weighted by the coupling $\kappa$. Secondly, the escort or coupled-probability $P_i^{(\kappa)} \equiv p_i^{1-\kappa} \Big/ \sum_{j=1}^n p_j^{1-\kappa}$ can be expressed as measuring the probability of a coupling of the states of system. This is demonstrated by multiplying the numerator and denominator by $\prod_{k=1}^n p_k^\kappa$

$$P_i^{(\kappa)} = \frac{p_i^{1-\kappa}}{\sum_{j=1}^n p_j^{1-\kappa}} = \frac{p_i \prod_{\substack{k=1 \\ k \neq i}}^n p_k^\kappa}{\sum_{j=1}^n \left( p_j \prod_{\substack{k=1 \\ k \neq j}}^n p_k^\kappa \right)} . \tag{2.1}$$

The expression on the right represents the normalized probability of the independent occurrence of event $i$ and $\kappa$ occurrences of the all the other events.

The definition for the coupled-exponential $\exp_\kappa(x) \equiv (1 + \kappa x)_+^{1/\kappa}$; $(a)_+ \equiv \max(0, a)$ when applied to multivariate distributions has the following difficulties. The argument of the coupled-Gaussian, $x = \dfrac{-y^2}{(2 + \kappa)\sigma_\kappa^2}$, depends on both the coupling itself and the coupled-variance, $\sigma_\kappa^2 = \dfrac{\int_{-\infty}^{\infty} x^2 f^{1-\kappa}(x)\, dx}{\int_{-\infty}^{\infty} f^{1-\kappa}(x)\, dx}$, where $\int_{-\infty}^{\infty} f^{1-\kappa}(x)\, dx \Big/ \int_{-\infty}^{\infty} f^{1-\kappa}(x)\, dx$ is the continuous form of the coupled-probability. A definition which allowed the argument to be independent of the coupling would better express the generalization. Secondly, for multiple dimensions the coupling is a function of the number of dimensions. Because of the dependence on the dimensions, the coupled-product $x \otimes_\kappa y \equiv (x^\kappa + y^\kappa - 1)^{1/\kappa}$ cannot be used to factor a multivariate distribution into the appropriate marginal distributions, since the marginals have different dimensions and hence different coupling values. These issues are addressed by defining the coupled-exponential such that the coupling term multiplying the argument and the coupling term forming the power are not necessarily equal.

**Definition 1  Coupled-Exponential:**  Given a variable $x$, the nonlinear statistical coupling $\kappa \in \mathbb{R}$ and an additional coefficient $a \in \mathbb{R}$ the *coupled-exponential* is defined as

$$\mathrm{e}^x_{\kappa_a} \equiv \exp_{\kappa_a}(x) \equiv \exp_\kappa(x;a) \equiv (1+\kappa x)^{\frac{1}{\kappa_a}}_+ \tag{2.2}$$

$$\kappa_a \equiv \left(\frac{1}{\kappa}+a\right)^{-1} \tag{2.3}$$

where $(y)_+ \equiv \max(0,\mathrm{y}) \forall y \in R.$

For finite $a$, $\lim_{\kappa \to 0} e^x_{\kappa_a} = e^x$ and for $a=0$, $\mathrm{e}^x_{\kappa_0} = \mathrm{e}^x_\kappa = (1+\kappa x)^{1/\kappa}_+$ which is the original definition of the coupled-exponential.  This more general definition will facilitate development of a generalized algebra for coupled random variables which are multivariate.  The inverse of this function is the coupled-logarithm.

**Definition 2  Coupled-Logarithm:**  Given a variable $x>0$ and nonlinear statistical coupling $\kappa \in \mathbb{R}$, the *coupled-logarithm* is defined as

$$\ln_{\kappa_a} x \equiv \ln_\kappa(x;a) \equiv \frac{1}{\kappa}\left(x^{\kappa_a}-1\right) \tag{2.4}$$

Anticipating the connection with dimensions, the coupled-product is defined such that the output coupling parameter is indexed by the sum of the input indices.  The definition is expressed using functions as the input, which later will be explicitly multidimensional distributions, but can be any scalar.

**Definition 3 Multivariate Coupled-Product**  Given two functions $f_{1,2}$ and a coupling $\kappa$ their coupled-product $\otimes_{\left(\kappa_{a_1},\kappa_{a_2}\right)}$ is

$$f_1 \otimes_{\left(\kappa_{a_1},\kappa_{a_2}\right)} f_2 \equiv \left(f_1^{\kappa_{a_1}}+f_2^{\kappa_{a_2}}-1\right)^{1/\kappa_{a_1+a_2}}_+ \tag{2.5}$$

The extension of the coupled-product to multiple functions is given by

$$\prod_{i=1}^n {}_{\otimes_{\kappa_{a_i}}} f_i \equiv \left(\sum_{i=1}^n f^{\kappa_{a_i}}-(n-1)\right)^{1/\kappa_{\Sigma a_i}} \tag{2.6}$$

The commutative and associative properties require that the coupling parameters follow the operands.  So the commutative property is

$$f_1 \otimes_{\left(\kappa_{a_1},\kappa_{a_2}\right)} f_2 = f_2 \otimes_{\left(\kappa_{a_2},\kappa_{a_1}\right)} f_1 \tag{2.7}$$

and the associative property is

$$\begin{aligned}&\left(f_1 \otimes_{\left(\kappa_{a_1},\kappa_{a_2}\right)} f_2\right) \otimes_{\left(\kappa_{a_1+a_2},\kappa_{a_3}\right)} f_3 =\\ &f_1 \otimes_{\left(\kappa_{a_1},\kappa_{a_2+a_3}\right)} \left(f_2 \otimes_{\left(\kappa_{a_2},\kappa_{a_3}\right)} f_3\right).\end{aligned} \tag{2.8}$$

The identity property requires a coupling index of zero operating on the one, $f_1 \otimes_{(\kappa_{d_1}, \kappa_0)} 1 = f_1$. The **coupled-division** function, subtracts the coupling indices

$$f_1 \oslash_{(\kappa_{d_1}, \kappa_{d_2})} f_2 \equiv \left( f_1 (\mathbf{x}_1)^{\kappa_{d_1}} - f_2^{\kappa_{d_2}} + 1 \right)_+^{1/\kappa_{d_1 - d_2}}. \qquad (2.9)$$

**Lemma 1** If $n$ operands $f_i(\mathbf{x}_i) = \exp_{\kappa_{d_i}} \left( \sum_{j=1}^{d_i} x_{i,j} \right)$, are multivariate coupled-exponentials of dimension $d_i$ then the output of the coupled-product operator is a multivariate coupled-exponential with dimension $D = \sum_{i=1}^{n} d_i$ and argument equal to the sum of the operand arguments, $\sum_{i=1}^{n} \sum_{j=1}^{d_i} x_{i,j}$.

*Proof.* Let $f_i(\mathbf{x}_i) = \exp_{\kappa_{d_i}} \left( \sum_{j=1}^{d_i} x_{i,j} \right)$, then applying Definition 3

$$\prod_{i=1}^{n} \otimes_{\kappa_{d_i}} f_i(\mathbf{x}_i) \equiv \left( \sum_{i=1}^{n} \left( 1 + \kappa \sum_{j=1}^{d_i} x_{i,j} \right)^{\frac{\kappa_{d_i}}{\kappa_{d_i}}} - (n-1) \right)^{\left( \frac{1}{\kappa} + \sum_{i=1}^{n} d_i \right)}$$

$$= \left( 1 + \kappa \sum_{i=1}^{n} \sum_{j=1}^{d_i} x_{i,j} \right)^{1/\kappa_D} = \exp_{\kappa_D} \left( \sum_{i=1}^{n} \sum_{j=1}^{d_i} x_{i,j} \right), \qquad (2.10)$$

which is a coupled-exponential function of dimension $D$ with the sum of the operand arguments. $\square$

The sum of coupled-logarithms is equivalent to the coupled-product of the arguments

$$\ln_{\kappa_a} A + \ln_{\kappa_b} B = \ln_{\kappa_{a+b}} \left( A \otimes_{(\kappa_a, \kappa_b)} B \right)$$

$$\tfrac{1}{\kappa} \left[ A^{\kappa_a} + B^{\kappa_b} - 2 \right] = \tfrac{1}{\kappa} \left[ \left( A^{\kappa_a} + B^{\kappa_b} - 1 \right)^{\frac{\kappa_{a+b}}{\kappa_{a+b}}} - 1 \right]. \qquad (2.11)$$

A coupled-exponential raised to a power is equivalent to

$$\left( e_{\kappa_d}^x \right)^a = e_{\kappa_d/a}^{ax}$$

$$= \left( 1 + \tfrac{\kappa}{a} (ax) \right)^{\frac{a(1+d\kappa)}{\kappa}}$$

$$= \left( (1 + \kappa x)^{\frac{(1+d\kappa)}{\kappa}} \right)^a. \qquad (2.12)$$

Note that the expression $\kappa_d / a$ modifies both $\kappa_d$ and $\kappa$; and can be expressed equivalently as $\dfrac{\kappa / a}{1 + d\kappa} = \dfrac{\kappa / a}{\left(1 + ad\left(\kappa / a\right)\right)} = \left(\kappa / a\right)_{ad}$. This relationship is used to define the generalization of the Gaussian

$$\left(e_{\kappa_d^{x^2}}\right)^{-\frac{1}{2}} = e_{-2\kappa_d}^{-x^2/2} \ . \tag{2.13}$$

This definition allows the coupling for a heavy tail distribution to define the source of nonlinearity from $0 < \kappa < \infty$ and $\infty > \kappa_d^{-1} > d$ to define the asymptotic fractal dimension of the distribution. Raising the argument of the coupled-logarithm to a power is equivalent to

$$\begin{aligned}
\ln_{\kappa_d} x^a &= a\ln_{a\kappa_d} x \\
&= \frac{a}{a\kappa}\left(x^{a\kappa_d} - 1\right) \\
&= \frac{1}{\kappa}\left(\left(x^a\right)^{\kappa_d} - 1\right)
\end{aligned} \tag{2.14}$$

The coupled-sum $x \oplus_\kappa y = x + y + \kappa xy$ does not bear directly on the factoring of the coupled-exponential, but its properties are summarized before proceeding. The coupled-sum is unchanged from the original algebra of nonextensive statistical mechanics, since it operates on the non-indexed coupling parameter. The coupled-sum of coupled-logarithms with the same coupling parameters is

$$\begin{aligned}
\ln_{\kappa_a} x \oplus_\kappa \ln_{\kappa_a} y &= \left(\frac{1}{\kappa}\left(x^{\kappa_a} - 1\right) \oplus_\kappa \frac{1}{\kappa}\left(y^{\kappa_a} - 1\right)\right) \\
&= \frac{1}{\kappa}\left(x^{\kappa_a} + y^{\kappa_a} - 2 + \left(x^{\kappa_a} - 1\right)\left(y^{\kappa_a} - 1\right)\right) \\
&= \ln_{\kappa_a}\left(xy\right).
\end{aligned} \tag{2.15}$$

Likewise, multiplication of coupled-exponentials results in the coupled-sum of the arguments $e_{\kappa_a}^x e_{\kappa_a}^y = e_{\kappa_a}^{x \oplus_\kappa y}$ . The distinction between the two coupling parameters $\kappa, \kappa_a$ , provides further clarification why the coupled-sum and coupled-product form a pseudo-algebra without a distributive property. These functions operate on distinct coupling parameters. Starting with the coupled-sum, a complete algebra with the distributive property has been defined by considering a deformation of the real numbers [9-11]. A separate algebra is formed by treating the coupled-product as a generalized sum and deriving a generalized multiplication which has a distributive property.

## 3. The multivariate coupled-Gaussian distribution

### 3.1. Definition of the multivariate coupled-Gaussian

The coupled-Gaussian is defined here such that the coupling is specified separately from the dimensions and the constant due to the power of 2 of the argument. Thus the distribution is defined with an exponent of $-1/2\kappa_d$. The parameters of the distribution are the coupled-mean $\boldsymbol{\mu}_{-2\kappa_d}$, coupled-variance $\boldsymbol{\Sigma}_{-2\kappa_d}$, and the coupled-partition function $Z_{-2\kappa_d}$. In the remainder of the text the subscript $-2\kappa_d$ will not be included and assumed to match the coupled-exponential. The multivariate coupled-Gaussian is defined as

$$G_{-2\kappa_d}(\mathbf{x};\boldsymbol{\mu},\boldsymbol{\Sigma}) \equiv \frac{1}{Z(\boldsymbol{\Sigma})}\exp_{-2\kappa_d}\left[-\tfrac{1}{2}(\mathbf{x}-\boldsymbol{\mu})^\top \cdot \boldsymbol{\Sigma}^{-1} \cdot (\mathbf{x}-\boldsymbol{\mu})\right]$$

$$= \frac{1}{Z(\boldsymbol{\Sigma})}\left(1+\kappa(\mathbf{x}-\boldsymbol{\mu})^\top \cdot \boldsymbol{\Sigma}^{-1} \cdot (\mathbf{x}-\boldsymbol{\mu})\right)_+^{\frac{1}{-2\kappa_d}}. \tag{2.16}$$

For $0<\kappa<\infty$ and equivalently $0>-2\kappa_d>-\frac{2}{d}$ the distribution is heavy-tail with larger values of coupling decreasing the rate of decay of the distribution. The multivariate Gaussian is approached in the limit as $\kappa \to 0$. For $-1/d<\kappa<0$ and equivalently $\infty>-2\kappa_d>0$ the distribution is compact-support and approaches the uniform multivariate distribution as $\kappa \to -1/d$. Coupling values less than $-1/d$ do not define a distribution. This definition of the multivariate coupled-Gaussian is identical to the multivariate t distribution with the coupling equal to the inverse of the degree of freedom, $\kappa = 1/\nu$.

The relationship between this definition of the multivariate coupled-Gaussian and the form using $q$-statistics is clarified by comparing equation (2.16) with the multivariate $q$-Gaussian defined by Umarov et. al. [12]

$$G_q(\mathbf{x};\boldsymbol{\mu},\boldsymbol{\Sigma}) \propto \exp_q\left[-\beta(\mathbf{x}-\boldsymbol{\mu})^\top \cdot \boldsymbol{\Sigma}^{-1} \cdot (\mathbf{x}-\boldsymbol{\mu})\right]$$

$$\propto \left(1-(1-q)\beta(\mathbf{x}-\boldsymbol{\mu})^\top \cdot \boldsymbol{\Sigma}^{-1} \cdot (\mathbf{x}-\boldsymbol{\mu})\right)^{\frac{1}{1-q}}. \tag{2.17}$$

In order for $\boldsymbol{\Sigma}$ to represent the generalized covariance the term multiplying the argument must be

$$(1-q)\beta = \left(\frac{2}{1-q}+d\right)^{-1} \equiv \frac{1-q_d}{2}. \tag{2.18}$$

Equating the exponent and arguments respectively for equation (2.17) and (2.16) leads to $\kappa_d = \frac{(1-q)}{-2}$ and $\kappa = \frac{(1-q_d)}{-2}$, with the understanding that the subscripted parameters for $\kappa$ and $q$ are defined separately in equations (2.3) and (2.18). There are in fact three translations which contribute to the new definition; $1-q \to \kappa$, which centers the Gaussian distribution at the origin, $\kappa \to \kappa_d$, which models the dependence of the exponent on the dimension, and $\kappa_d \to -2\kappa_d$, which makes $\kappa$ positive for heavy-tail distributions and simplifies its relationship to the degree of freedom and other sources of nonlinearity. Accounting for the effect of the negative sign the shift in parameters between the argument and the exponent can also be understood to be $-2\kappa \to 1/2(-\kappa)_{-d}$ and equivalently $1-q_d \to 1/(1-q)$, thus in both cases the subscript for the exponent is reduced by $d$.

The partition function is determined by the normalization

$$Z[\boldsymbol{\Sigma}] \equiv \int_{R^d} \exp_{-2\kappa_d}\left[-\tfrac{1}{2}(\mathbf{x}-\boldsymbol{\mu})^{\top}\cdot\boldsymbol{\Sigma}^{-1}\cdot(\mathbf{x}-\boldsymbol{\mu})\right]d\mathbf{x}. \qquad (2.19)$$

Following the methods in [12] the integral can be expressed as a one-dimensional integral using the radial symmetry. A change of variables $\mathbf{y} = \mathbf{x}/\sqrt{2^d|\boldsymbol{\Sigma}|}$, accounts for the covariance. The surface area of the unit circle in $d$ dimensions is $d\pi^{d/2}/(d/2)!$, where a generalized factorial is used instead of the gamma function $x! \equiv \Gamma(x+1)$. And the radial integral is $\int_0^\infty r^{d-1}e_{-2\kappa_d}^{-r^2}dr$ giving

$$Z[\Sigma] = \sqrt{2^d|\boldsymbol{\Sigma}|}\,\frac{d\pi^{d/2}}{(d/2)!}\int_0^\infty r^{d-1}e_{-2\kappa_d}^{-r^2}dr \qquad (2.20)$$

for the partition function, whose solution is

$$Z[\boldsymbol{\Sigma}] = \sqrt{(2\pi)^d|\boldsymbol{\Sigma}|}\begin{cases} \dfrac{(1+d\kappa)\,\frac{1}{2\kappa}!}{\sqrt{(2\kappa)^d}\,\frac{1}{2\kappa_d}!} & \kappa > 0 & \text{Heavy-tail} \\[3mm] 1 & \kappa = 0 & \text{Gaussian} \\[3mm] \dfrac{1}{\sqrt{(-2\kappa)^d}}\,\dfrac{\frac{-1}{2\kappa_d}!}{\frac{-1}{2\kappa}!} & -1\!\!\Big/\!d < \kappa < 0 & \begin{array}{l}\text{Compact-}\\\text{support.}\end{array} \end{cases} \qquad (2.21)$$

The coupled-mean and covariance parameters of the distribution [13] are defined using the coupled-probability; so the coupled-mean after canceling out the normalization is

$$\mu_i \equiv \frac{\int_{R^d} x_i\left[\exp_{-2\kappa_d}\left[-\tfrac{1}{2}(\mathbf{x}-\boldsymbol{\mu})^{\top}\cdot\boldsymbol{\Sigma}^{-1}\cdot(\mathbf{x}-\boldsymbol{\mu})\right]\right]^{1+2\kappa_d}d\mathbf{x}}{\int_{R^d}\left[\exp_{-2\kappa_d}\left[-\tfrac{1}{2}(\mathbf{x}-\boldsymbol{\mu})^{\top}\cdot\boldsymbol{\Sigma}^{-1}\cdot(\mathbf{x}-\boldsymbol{\mu})\right]\right]^{1+2\kappa_d}d\mathbf{x}} \qquad (2.22)$$

and the coupled-covariance is

$$\Sigma_{ij} \equiv \frac{\int_{R^d} x_i x_j\left[\exp_{-2\kappa_d}\left[-\tfrac{1}{2}(\mathbf{x}-\boldsymbol{\mu})^{\top}\cdot\boldsymbol{\Sigma}^{-1}\cdot(\mathbf{x}-\boldsymbol{\mu})\right]\right]^{1+2\kappa_d}d\mathbf{x}}{\int_{R^d}\left[\exp_{-2\kappa_d}\left[-\tfrac{1}{2}(\mathbf{x}-\boldsymbol{\mu})^{\top}\cdot\boldsymbol{\Sigma}^{-1}\cdot(\mathbf{x}-\boldsymbol{\mu})\right]\right]^{1+2\kappa_d}d\mathbf{x}}. \qquad (2.23)$$

### 3.2. Constructing the multivariate coupled-Gaussian from univariates

The multiplication of univariate Gaussians forms a multivariate Gaussian with a diagonal covariance matrix, reflective of the independence between the variables. The multivariate coupled-Gaussian with a diagonal generalized covariance, can now be constructed from its univariate distributions using the newly defined coupled-product. The dependency can be described as a coupling between the variates. A caveat is that the normalization must be treated separately.

**Lemma 2** Given $d$ unnormalized univariate coupled-Gaussian distributions

$$f\left(x_i; \mu_i, \sigma_i^2\right) = \exp_{-2\kappa_1}\left[-\frac{\left(x_i - \mu_i\right)^2}{2\sigma_i^2}\right] \text{ with } \kappa > -\frac{1}{d}, \text{ the coupled-product with } -2\kappa_1$$

results in the unnormalized multivariate Gaussian of $d$ dimensions,

$$g\left(\mathbf{x}; \boldsymbol{\mu}, \boldsymbol{\Sigma}\right) = \exp_{-2\kappa_d}\left[-\tfrac{1}{2}\left(\mathbf{x} - \boldsymbol{\mu}\right)^{\mathsf{T}} \boldsymbol{\Sigma}^{-1}\left(\mathbf{x} - \boldsymbol{\mu}\right)\right] \text{ with } \boldsymbol{\Sigma} = \begin{pmatrix} \sigma_1^2 & 0 & \dots & 0 \\ 0 & \sigma_2^2 & \dots & 0 \\ \dots & \dots & \dots & \dots \\ 0 & 0 & \dots & \sigma_d^2 \end{pmatrix} \text{ and}$$

$$\boldsymbol{\mu} = \left(\mu_1, \mu_2, \dots, \mu_d\right)^{\mathsf{T}}.$$

*Proof.* Applying Lemma 1, the coupled product of the $d$ functions $f$ is

$$\prod_{i=1}^{d}{}_{\otimes_{-2\kappa_1}} \exp_{-2\kappa_1}\left[-\frac{\left(x_i - \mu_i\right)^2}{2\sigma_i^2}\right] = \exp_{-2\kappa_d}\left[-\tfrac{1}{2}\sum_{i=1}^{d}\frac{\left(x_i - \mu_i\right)^2}{\sigma_i^2}\right]. \tag{2.24}$$

and $\sum_{i=1}^{d}\dfrac{\left(x_i - \mu_i\right)^2}{\sigma_i^2} = \left(\mathbf{x} - \boldsymbol{\mu}\right)^{\mathsf{T}} \boldsymbol{\Sigma}^{-1}\left(\mathbf{x} - \boldsymbol{\mu}\right)$ with $\boldsymbol{\Sigma} = \begin{pmatrix} \sigma_1^2 & 0 & \dots & 0 \\ 0 & \sigma_2^2 & \dots & 0 \\ \dots & \dots & \dots & \dots \\ 0 & 0 & \dots & \sigma_d^2 \end{pmatrix}$ and

$\boldsymbol{\mu} = \left(\mu_1, \mu_2, \dots, \mu_d\right)^{\mathsf{T}}. \ \square$

For a set of normalized coupled Gaussians, the procedure for forming the normalized multivariate coupled Gaussian is to first multiply each input distribution by its partition function, $Z$, combine the distributions using the newly defined coupled product, and finally normalize the result with the its respective partition function. This issue is not simplified by using the canonical form, $\exp_{-2\kappa_d}\left[\boldsymbol{\theta}^1\mathbf{x}_1 + \boldsymbol{\theta}^2\mathbf{x}_2 - \psi[\boldsymbol{\theta}^1, \boldsymbol{\theta}^2]\right]$ because each of the parameters $\boldsymbol{\theta}^1, \boldsymbol{\theta}^2, \psi$ depends on the partition function, as shown in [14]. An interesting research question is whether an alternative constraint than normalization, would result in a definition for the canonical parameters which are additive when forming the multivariate from marginal distributions.

The domain of the univariate coupled-Gaussians being combined by the coupled-product is $\kappa > -\frac{1}{d}$. This insures that the multivariate distribution has a decaying tail. A question for exploration is why the compact-support distributions in the domain $-1 < \kappa < -1/d$ cannot be coupled to form a $d$-dimensional distribution.

The ability to properly form multivariate distributions from the coupled-product of marginal distributions is important for the generalization of probabilistic graphs [15], which are based on identification of conditionally independent components. It is anticipated that such a construction will enable a generalization of Markov and Bayesian networks, in which the fusion of variables is based on the coupled-product defined by equation (2.6). This has the potential of simplifying the network topology for models requiring complex dependencies. A complete definition of such a network will require application of the coupled-division and consideration of conditional distributions.

## 4. Conclusion

The proposed definition for the coupled-exponential isolates the nonlinear statistical coupling parameter $\kappa$ (or simply the coupling) from the argument of the coupled-exponential distributions, and the dimensionality and power of the argument. The power of the coupled-logarithm and the inverse power of the coupled-exponential for a $d$-dimensional argument is $\kappa_d \equiv \left(\frac{1}{\kappa} + d\right)^{-1}$. This term is multiplied by $-1$ for the coupled-exponential distribution and $-2$ for the coupled-Gaussian distribution; and the negative sign makes the heavy-tail domain $0 < \kappa < \infty$. The resulting multivariate coupled-Gaussian distribution, equation (2.16) is identical to the multivariate t-distribution with the coupling parameter equal to the inverse of the degree of freedom. This simple relationship is anticipated to also be true of other properties of complex systems, enabling further clarification of the physical interpretation of nonlinear statistical coupling.

The coupling between distributions is defined using a multivariate form of the coupled product, equation (2.6), which sums the arguments of coupled-exponentials and sums the dimension which indexes the coupling, $\kappa_D \equiv \left(\frac{1}{\kappa} + D\right)^{-1}, D = \sum_{i=1}^{n} d_i$. This is required because the rate of the decay of a coupled-exponential depends on the number of dimensions. Isolating this factor, is needed to develop models of physical processes such as generalized Markov random fields based on coupled-Gaussians.

**Acknowledgement**

Discussions with Brian Scannell regarding applications of the coupled algebra to recursive Bayesian computations highlighted the need to account for dimensionality in the use of the coupled-product to form multivariate distributions. The research was supported in part by a Raytheon IDEA project titled Coupled Bayesian Networks.